# Imaging nanoscale Fermi surface variations in an inhomogeneous superconductor

W. D. Wise[1], Kamalesh Chatterjee[1], M. C. Boyer[1], Takeshi Kondo[2,1*], T. Takeuchi[2,3], H. Ikuta[2], Zhijun Xu[4], Jinsheng Wen[4], G. D. Gu[4], Yayu Wang[1†] & E. W. Hudson[1]

[1]*Department of Physics, Massachusetts Institute of Technology, Cambridge, MA 02139, USA.*
[2]*Department of Crystalline Materials Science, Nagoya University, Nagoya 464-8603, Japan.*
[3]*EcoTopia Science Institute, Nagoya University, Nagoya 464-8603, Japan.*
[4]*Condensed Matter Physics and Materials Sciences Department, Brookhaven National Lab, Upton, NY 11973, USA.*
*Present address: Ames Laboratory and Dept. of Physics and Astronomy, Iowa State University, Ames, IA 50011
†Present address: Department of Physics, Tsinghua University, Beijing 100084, China.

**Particle-wave duality suggests we think of electrons as waves stretched across a sample, with wavevector *k* proportional to their momentum. Their arrangement in "*k*-space," and in particular the shape of the Fermi surface, where the highest energy electrons of the system reside, determine many material properties. Here we use a novel extension of Fourier transform scanning tunneling microscopy to probe the Fermi surface of the strongly inhomogeneous Bi-based cuprate superconductors. Surprisingly, we find that rather than being globally defined, the Fermi surface changes on nanometer length scales. Just as shifting tide lines expose variations of water height, changing Fermi surfaces indicate strong local doping variations. This discovery, unprecedented in any material, paves the way for an understanding of other inhomogeneous characteristics of the cuprates, like the pseudogap magnitude, and highlights a new approach to the study of nanoscale inhomogeneity in general.**

That high temperature superconductors should exhibit nanoscale inhomogeneity is unsurprising. In correlated electron materials, Coulomb repulsion between electrons hinders the formation of a homogeneous Fermi liquid, and complex real space phase separation is ubiquitous[1]. Scanning tunneling microscopy (STM) measurements have revealed significant spectral variations in a number of cuprates including $Bi_2Sr_2CuO_{6+x}$ (Bi-2201)[2] and $Bi_2Sr_2CaCu_2O_{8+x}$ (Bi-2212)[3-5].

This intrinsic inhomogeneity poses challenges to the interpretation of bulk or spatially averaged measurements. For example, angle resolved photoemission spectroscopy (ARPES) is a powerful technique for studying *k*-space structure in the cuprates[6]. However, ARPES can only provide spatially-averaged results, and uniting these with the nanoscale disordered electronic structure measured by STM remains a formidable task.

Our approach to addressing this issue originates from discoveries by Fourier transform scanning tunneling microscopy (FT-STM), which has emerged as an important tool for studying the cuprates. These studies begin with the collection of a spectral survey, in which differential conductance spectra, proportional to local density of states (LDOS), are measured at a dense array of locations, creating a three dimensional dataset of LDOS as a function of energy and position in the plane. By Fourier transforming constant energy slices of these surveys, referred to as LDOS or conductance maps, FT-STM allows the study of two phenomena linked to the cuprate FS (Fig. 1b). First, non-dispersive wavevectors of the checkerboard-like charge order observed in many cuprates[7-10] are likely connected to the FS-nesting wavevectors near the anti-nodal $(\pi,0)$ Brillouin zone boundary (e.g. arrow in Fig. 1b)[11]. Second, dispersive quasiparticle interference (QPI) patterns[12-14] originate from elastic scattering of quasiparticles on the Fermi surface near the nodal $(\pi, \pi)$ direction[15]. Taken together, these phenomena provide complementary information about the cuprate FS.

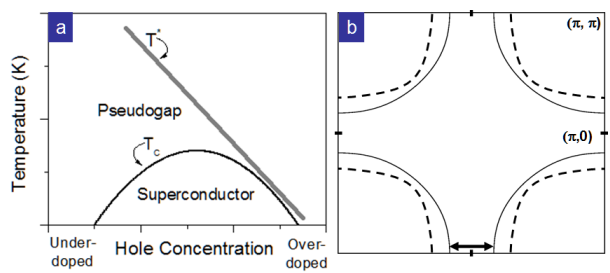

Figure 1. **Phase diagram and Fermi surface topology of the cuprates**. (a) A minimal generic phase diagram of the high temperature superconductors shows a superconducting transition temperature $T_C$ which is parabolic with doping, peaking at optimal doping, while the pseudogap temperature $T^*$ – and the proportional pseudogap magnitude $\Delta_{PG}$ – decrease nearly linearly with doping. (b) The hole-doped cuprate Fermi surface is typically seen as hole-like, closing around empty states centered at $(\pi,\pi)$, rather than filled states centered at $(0,0)$. Moving from optimally doped (solid line) to underdoped (dashed) materials, the hole pockets shrink. This increases the length of the nesting vector (arrow) near the anti-node.

However, because these phenomena were previously characterized using Fourier transforms of large LDOS maps containing a wide range of energy gaps and spectra, previous FT-STM mapping of the FS was still spatially-averaged.[16] The atomic scale spatial resolution of STM was not exploited, so connections between FS geometry and local electronic structure went unexamined.

Here we introduce two new STM analysis techniques which allow extraction of a *local* FS. In studies of Bi-2201 and Bi-2212, we find that the cuprate FS varies at the nanometer scale, and that its local geometry correlates strongly with the size of the large, inhomogeneous energy gap that has been extensively studied by STM[4] and which we associate with the pseudogap[2].

We first investigate the spatial dependence of the anti-nodal FS using checkerboard charge order. Our recent study of Bi-2201 showed that the average checkerboard wavevector decreases with increased doping[11]. This trend, inconsistent with many proposed explanations of the checkerboard, matches the doping dependence of the anti-nodal FS-nesting wavevector (Fig. 1b), and led us to conclude that the checkerboard is caused by a FS-nesting induced charge density wave. Here we continue the investigation of the three Bi-2201 dopings considered in our previous work, two underdoped with superconducting transitions at 25 K (UD25) and 32 K (UD32), and one optimally doped with $T_C$ = 35 K (OP35). Although, following convention, we previously reported FT-measured, spatially averaged wavevectors, careful observation of the checkerboard pattern (Fig. 2a) shows that the periodicity changes drastically with position.

One way of analyzing this variation is with a Voronoi diagram. After identifying local peaks of the checkerboard modulation (red dots in Fig. 2a), we divide the map into cells, each containing points closer to one checkerboard maximum than any other. The square root of the cell size is a measure of the local checkerboard wavelength. We find that this local wavelength is highly correlated with the previously observed gap size inhomogeneity (Fig. 2b,c), with a correlation coefficient of -0.4.

Another method of investigating this relationship between local checkerboard periodicity and gap size is to modify the traditional FT technique by first masking the LDOS map by gap size. This technique is illustrated in Fig. 3. The LDOS map is set to zero everywhere outside a desired gap range and then Fourier transformed to reveal a gap dependent checkerboard wavevector. The result is qualitatively similar to the Fourier transform of the complete map, but the wavevector measured is due solely to the fraction of the sample within the selected range of gap magnitudes. Fourier transforms of different regions reveal different wavevectors (Fig 3b, c). Consistent with previously reported sample averages,[11] wavevectors increase with gap size (Fig. 3d).

These two independent techniques not only demonstrate the inhomogeneity of the checkerboard wavelength, the likely cause of universally reported short checkerboard correlation lengths[11], but also reveal that the checkerboard wavevector and local gap size are strongly correlated. Between samples, average checkerboard wavevector decreases with increased doping[11], consistent with the decrease of the anti-nodal FS-nesting wavevector (Fig. 1b). The tunneling measured gap size (scaling with pseudogap temperature T* of Fig. 1a) also on average decreases with increased doping. Thus the positive correlation of local gap size and checkerboard wavevector is consistent with a picture in which local FS variations, driven by local doping variations, affect both. Notably, where gap sizes from different samples overlap, so do their checkerboard wavelengths (Fig. 3d), indicating that checkerboard properties are truly set by local rather than sample average properties.

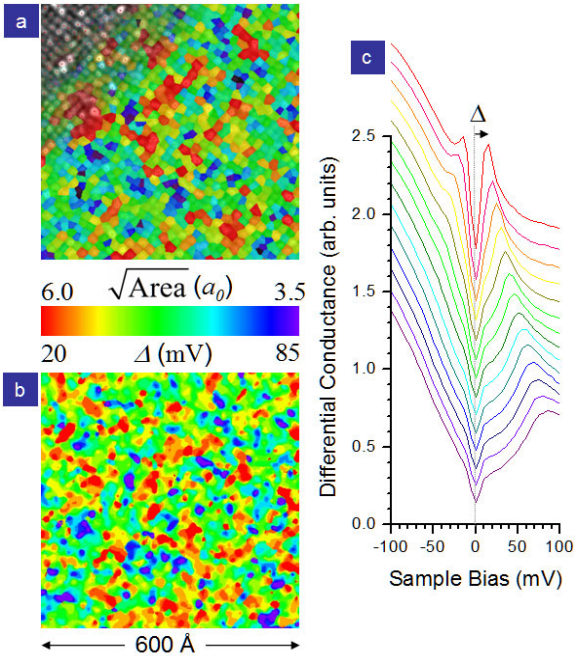

Figure 2. **Local variations of the Bi-2201 checkerboard**. (a) Conductance map (E = +10 mV slice of a 600 Å spectral survey) of $T_C$ = 32 K underdoped Bi-2201 shows a spatially varying checkerboard charge modulation (upper left). Voronoi cells, associated with checkerboard maxima (red dots) and colored to indicate their area, allow determination of local wavelength. (b) Traditional gap map of the same area shows well known variations of gap size $\Delta$. (c) Spectra from the survey (sorted, averaged and colored by gap size) highlight the remarkable low energy homogeneity in the presence of strong higher energy inhomogeneity.
Spectral survey parameters: $I_{set}$ = 400 pA, $V_{sample}$ = -200 mV.

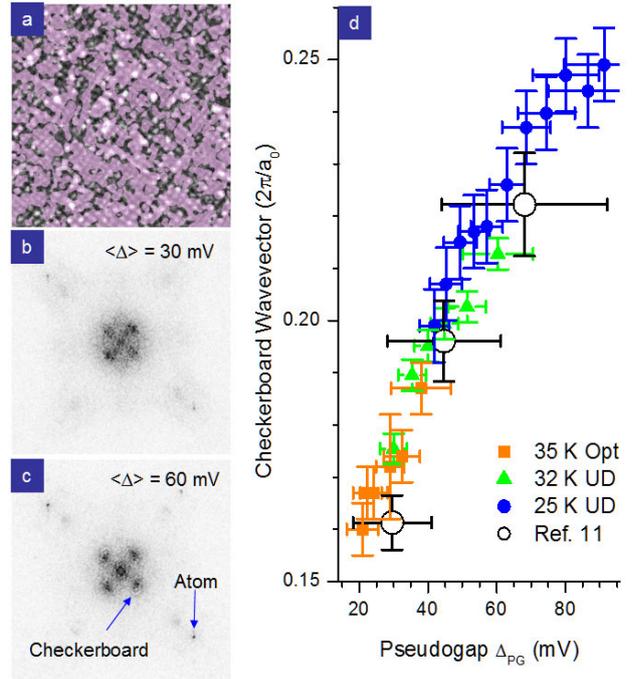

Figure 3. **Gap masked Fourier transform measurements of the Bi-2201 checkerboard**. (a) Using the gap map of Fig. 2b we mask the conductance map of Fig. 2a, zeroing out (shading pink) data with gaps outside a desired range, here $\Delta$ = 40 mV to 55 mV. (b,c) Fourier transforms of the masked data (with (b) $<\Delta>$ = 30 mV and (c) $<\Delta>$ = 60 mV) show checkerboard wavevectors, whose length can be compared to the atomic periodicity, that shift with gap masking range. (d) Checkerboard wavevectors for this sample (green triangles) as well as optimally doped 35 K (orange square) and underdoped 25 K (blue circles) samples. Overlaid are large area averages from our previous work.[11] Error bars indicate standard deviation of gap range used (horizontal) and FFT peak fit accuracy (vertical).

In order to further investigate this idea we next turn to quasiparticle interference (QPI) studies of slightly overdoped ($T_C$ = 89 K) Bi-2212. The idea behind QPI, pioneered by the Davis group[12-14] and Dung-Hai Lee,[15] is illustrated in Fig. 4a. Interference patterns arising from quasiparticle scattering are dominated by wavevectors connecting $k$-space points of high density of states. For any given energy, eight such symmetric points exist, all on the FS. The well defined wavevectors (colored lines) of the resultant interference pattern can therefore be used to reconstruct the Fermi surface.

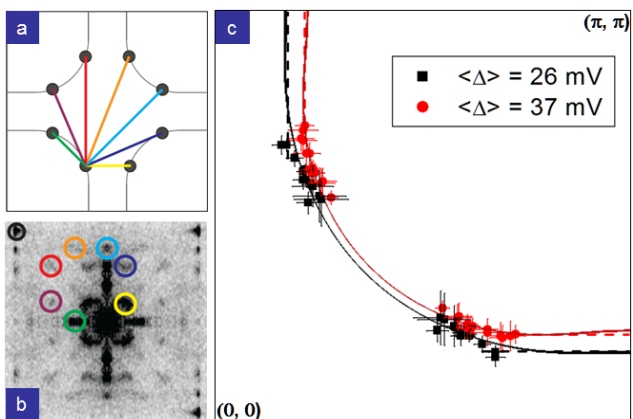

Figure 4. **Quasiparticle interference derived local FS in Bi-2212**. (a) A schematic of the FS (solid line) in the first Brillouin zone, shows symmetry, leading to an eight fold replication of any points at which the density of states peak (e.g. circles). Scattering between quasiparticles at these points lead to a set of interference wavevectors (arrows), corresponding to peaks in interference maps like (b) a Fourier transform of a 600 Å, 400 pixel, E = 12 mV conductance map of 89 K OD Bi-2212. The positions of these peaks (defined in terms of the atomic wavevector circled in black) uniquely define a position in k-space on the FS. Fitting interference peaks in a series of FFT maps at various energies from two different masks of the same data leads to (c), two different "local Fermi surfaces" (solid symbols with error bars indicating standard deviation for values obtained from different interference peaks). Dashed lines, obtained from checkerboard determined nesting wavevectors, extend the determined FS to the anti-node. Solid lines are Fermi surfaces from a rigid band, tight binding model[17] at two different dopings, p = 0.10 and 0.18.

Just as in checkerboard studies, previous work on QPI has yielded spatially averaged results[12-14]. As above, we extend QPI analysis to yield local, gap dependent information. The interference wavevectors (circled in Fig. 4b) found in Fourier transforms of gap masked conductance maps can be inverted to derive a FS[13], now associated with the gap range of the mask. Doing this for two different gap ranges, from 30 mV to 60 mV ($<\Delta>$ = 37 mV), and 10 mV to 30 mV ($<\Delta>$ = 26 mV), we find distinct shifts in the FS (Fig. 4c). We extend this to the anti-node with checkerboard order, resolvable in Bi-2212 as non-dispersive order at energies above where the QPI signal weakens. Adding in nested anti-nodal FS segments (dashed lines) derived from checkerboard periodicity, we arrive at a nearly complete view of different local Fermi surfaces corresponding to different spatial locations, correlated spatially with different gap sizes. We also plot in Fig. 4c rigid band tight binding Fermi surfaces[17] from two different dopings (p = 0.10 and 0.18 as calculated from the pocket area) very similar to the surfaces we derive.

Although $k$-space variation on nanometer length scales may at first glance seem shocking, upon further reflection this result is not entirely surprising. Raising or lowering a uniformly slanted sea floor near the shore (changing the amount of sea above the floor) changes the position of the shoreline. Analogously, raising or lowering local doping changes the local Fermi surface. McElroy et al. have even demonstrated that the locations of dopant oxygen atoms correlate with local gap size variations[18]. Interestingly, the correlation McElroy *et al.* found was the opposite of what one might at first expect from a local doping picture. While oxygen dopants contribute holes and hence increase the global doping of the sample, they correlate with an *increased* local gap size, consistent with underdoping. This led the authors and others[19] to declare that variation in gap size is unlikely to be charge driven and instead propose variations in local pairing-potential.

The results we report here, however, cannot be explained by pairing-potential inhomogeneity. Instead, local doping variations appear most consistent with our results. Although McElroy *et al.* suggest that these variations cannot be explained by hole accumulation models, Zhou *et al.* claim that they have missed the oxygen atoms responsible for inhomogeneity[20]. Alternatively, local doping variations could be driven by dopant generated strain. This would also explain the correlation of gap variations with the strain-associated structural supermodulation in Bi-2212[21].

Regardless of the exact cause of these local doping variations, they can explain several previous results. Perhaps the clearest examples consist of recent STM results from the Yazdani group showing that the gap closing temperature varies spatially, scaling with local gap size,[5] and that both are correlated with higher temperature electronic structure[22]. Those results are unsurprising given the local Fermi surface variations we report here.

Considering the nature of the Fermi surface variations in detail, we find that the locally determined Fermi surfaces converge near the nodes while they are strongly inhomogeneous in the anti-nodes (Fig. 4c). This could explain the ARPES-measured dichotomy of coherent nodal / incoherent anti-nodal quasiparticle excitations found in a variety of cuprates.[6,23] This differentiation is strongest in underdoped samples which could also arise from the inhomogeneity we report here. ARPES sums signal from differently doped regions; as more highly doped regions yield higher signal, ARPES will overemphasize them. Coupled with the observation that the width of the gap (and hence doping) distribution scales with mean gap size[24], and is thus smaller in overdoped than underdoped samples, inhomogeneity should have a stronger effect on ARPES measurements in underdoped than in overdoped samples. This effect is particularly apparent in Bi-2201, which is more inhomogeneous than Bi-2212.[2]

Despite the success of this interpretation, some outstanding questions remain. The model curves[17] of Fig. 4c suggest that the effective band energy may shift by as much as 20 mV between different regions of the sample. This shift would lead to strong scattering, even in the nodal direction. However, aside from the reasonable match to our extracted Fermi surfaces, there is no reason to believe that this global average-extracted rigid band model should completely describe the local Fermi surfaces. For example, our extracted Fermi surfaces appear to converge more quickly headed into the nodal region than the model surfaces.

Another question concerns a homogeneous gap we have reported.[2] The large, inhomogeneous gap discussed throughout this paper is probably more accurately termed the pseudogap, while we identified as the superconducting gap a second, relatively homogeneous smaller gap which opens at $T_C$. One might imagine that superconductivity, as characterized by size of the superconducting gap, should be as strongly affected by inhomogeneous local doping as the pseudogap. This is not what we have observed[2]. One explanation is that doping dependence differences make inhomogeneity affect the pseudogap more than the superconducting gap (the pseudogap, scaling with $T^*$, changes more than the superconducting gap, scaling with $T_C$). Another explanation may lie in their momentum space distribution. Raman spectroscopy[25] and ARPES[26,27] results indicate that the superconducting gap is most strongly associated with near-nodal states, while the pseudogap arises near the anti-nodes. As noted above, the nodal region is significantly more homogeneous than the antinodal, and hence could lead to more homogeneous superconducting than pseudogap properties.

This interpretation also points towards an explanation of bulk measurement results. Although several are suggestive of nanoscale inhomogeneity[4], including neutron measurements of the magnetic resonance peak width[28], thermodynamic measurements appear inconsistent with strong inhomogeneity[29]. These measurements, however, are most sensitive to the nature of the *superconducting* gap and the low energy density of states, both of which appear homogeneous. Undoubtedly these homogeneous properties relate to the homogeneity of the near nodal Fermi surface. Nonetheless, they are remarkable given the large local doping inhomogeneity and related Fermi surface variations we report here, and beg further experimental and theoretical investigation.

## Acknowledgements
We thank A.V. Balatsky, N. Gedik, J.E. Hoffman, K.M. Lang, P.A. Lee, Y. Lee, T. Senthil and Z. Wang for helpful comments. This research was supported in part by a Cottrell Scholarship awarded by the Research Corporation, by the MRSEC and CAREER programs of the NSF, and by DOE.

## Author Contributions
WDW, KC and MCB shared equal responsibility for most aspects of this project from instrument construction through data collection and manuscript preparation. WDW performed a majority of the analysis. TK grew the Bi-2201 samples and helped refine the STM. TT and HI contributed to Bi-2201 sample growth. ZX, JW and GDG contributed to Bi-2212 sample growth. YW contributed to analysis and manuscript preparation. EWH advised.

*Correspondence and requests for materials should be addressed to EWH (ehudson@mit.edu).*